\documentclass[aps,prc,twocolumn,showpacs,superscriptaddress,groupedaddress]{revtex4-1}  
\usepackage{mathrsfs,amsmath,amssymb,amsthm}
\usepackage{amssymb,fmtcount}
\usepackage{graphicx,epsfig,latexsym,overpic,amssymb,color}
\usepackage{threeparttable}
\preprint{\today}

\begin{document}
\title{
Bayesian evaluation of charge yields of fission fragments of $^{239}$U
}
\author{C.Y. Qiao}
\affiliation{
State Key Laboratory of Nuclear Physics and Technology, School of Physics,
Peking University, Beijing 100871, China
}
\author{J.C. Pei}\email{peij@pku.edu.cn}
\affiliation{
State Key Laboratory of Nuclear Physics and Technology, School of Physics,
Peking University, Beijing 100871, China
}
\author{Z.A. Wang}
\affiliation{
State Key Laboratory of Nuclear Physics and Technology, School of Physics,
Peking University, Beijing 100871, China
}
\author{Y. Qiang}
\affiliation{
State Key Laboratory of Nuclear Physics and Technology, School of Physics,
Peking University, Beijing 100871, China
}
\author{Y.J. Chen}
\affiliation{
China Institute of Atomic Energy, Beijing 102413, China
}
\author{N.C. Shu}
\affiliation{
China Institute of Atomic Energy, Beijing 102413, China
}
\author{Z.G. Ge}
\affiliation{
China Institute of Atomic Energy, Beijing 102413, China
}

\begin{abstract}
Recent experiments [Phys. Rev. Lett. 123, 092503(2019); Phys. Rev. Lett. 118, 222501 (2017)] have made remarkable progress in measurements of
the isotopic fission-fragment yields of the compound nucleus $^{239}$U, which is of great interests for fast-neutron reactors and for benchmarks of fission
models. We apply the Bayesian neural network (BNN) approach to
learn existing evaluated charge yields and infer the incomplete charge yields of $^{239}$U.
We found the two-layer BNN is improved compared to the single-layer BNN for the overall performance.
Our results support the normal charge yields of $^{239}$U around Sn and Mo isotopes.
The role of odd-even effects in charge yields has also been studied.

\end{abstract}
\maketitle
\section{introduction}

Nuclear fission is a very complex non-equilibrium quantum many-body dynamic process
and a deeper understanding of fission presents a well-known challenge in nuclear physics~\cite{Krappe2012}.
There are still strong motivations to study nuclear fission with increasing wide
nuclear applications~\cite{nd} in productions of energies and rare isotopes, and in fundamental physics
such as synthesizing superheavy elements~\cite{SHN2013,JCP2009PRL} and constraints on $r$-process~\cite{PRL111(2013)242502}.
In particular, the high-quality energy dependent fission data is very needed for fast-neutron reactors.
However, fission measurements are very difficult and fission data are generally incomplete
and have large uncertainties. In major nuclear data libraries~\cite{ENDF,JENDL,JEFF,CENDL},
evaluated fission yields are only available at thermal neutron energies, 0.5 and 14 MeV.

Recently, the isotopic $^{239}$U fission products with close excitation energies have been measured experimentally by different methods~\cite{PRL2017U,PRL2019U}.
Previously, the fission fragment mass distributions or charge distributions can be obtained. However, precise measurements
of full isotopic fission yields are only possible very recently with the inverse kinematics and magnetic spectrometers~\cite{PRC88(2013)024605,PRC91(2015)064616,PRC95(2017)054603,PRC97(2018)054612}.
The correlated fission observables are very crucial for deeper understandings of fission process. It was reported that
the charge yields around Sn and Mo are exceptionally small in the $^{238}$U(n, f) reaction~\cite{PRL2017U} but it was normal in
the later experiment on $^{239}$U via transfer reactions~\cite{PRL2019U}. It would be interesting to evaluate the two
discrepant experimental results.

The microscopic fission dynamical models based on potential energy surfaces (PES) can obtain
reasonable fission mass yields and charge yields~\cite{PRC93(2016)054611}.  In addition, the non-adiabatic time-dependent
density functional theory~\cite{PRC93(2016)011304} is promising to obtain various fission observables such as
fission yields, total kinetic energies and neutron multiplicities, due to developments of
supercomputing capabilities. The macro-microscopic fission models based on
complex PES have been successfully used for descriptions of fission yields~\cite{PRL106(2011)132503}. Generally fission models
with more predictive ability would have less precision.
For precise evaluations of
fission data, phenomenology models such as Brosa model~\cite{Brosa} and the recent GEF model~\cite{GEF} are very successful and have been widely used.

It is known that machine learning is powerful to learn and infer from complex big data,
which is of great interests in interdisciplinary physics subjects.
Recently, it was shown that Bayesian neural network can be used for evaluations of
incomplete
fission mass yields with uncertainty quantifications~\cite{fissionPRL2019}.
The machine learning has been used in nuclear physics such as  the extrapolation of nuclear masses~\cite{Utama2016PRC,Niu2018PLB,Neufcourt2018PRC},
 fission yields~\cite{fissionPRL2019,amy},  various nuclear structure~\cite{Niu2019PRC,bai,keeble,Lasseri,Utama2016JPG,jiang} and reaction observables~\cite{Ma2020CPC014104,Ma2020CPC124107,amy2}.
The machine learning has also been widely applied in other physics subjects, such as the constrains of equation of state of neutron stars from gravitational wave signals~\cite{gw}
and for facilitating the lattice QCD calculations~\cite{qcd}.
We speculate that machine learning is promising for developing new evaluation methods of nuclear data, in regarding to correlated fission observables and existing large uncertainties.

Previously, we have applied BNN to evaluate fission mass yields~\cite{fissionPRL2019}.
In this work, we apply BNN to evaluate the fission charge yields, in particular
the discrepant  charge yields of the compound nucleus $^{239}$U.
The charge distribution data is usually scare and is very useful in nuclear applications.
There could be different energy dependent behaviors of charge distributions and mass distributions~\cite{charge}.
The charge distributions show distinct odd-even effects~\cite{charge2}, while odd-even effects are ambiguous in mass distributions.
Compared to our previous work, we employ a multi-layer neural network in this work.

\section{The Models}

The BNN approach~\cite{Neal1996} adopts probability distributions as connection weights and is naturally suitable for uncertainty quantifications,
in contrast to standard neural networks which optimize definite values for connection weights.
The BNN approach to statistical inference is based on Bayes' theorem, which provides a connection between a given hypothesis(in terms of  problem-specific beliefs for a set of parameters $\omega$) and a set of data ($x,t$) to a posterior probability \emph{p}($\omega$$\mid$x,t) that is used to make predictions on new inputs, which is written as
\begin{equation}
    p(\omega | x,t)=\frac{p(x,t| \omega)p(\omega)}{p(x,t)},
                                                                  \label{eqn.01}
\end{equation}
where \emph{p}(x,t$\mid$$\omega$)is the `likelihood' that a given model describes the data and \emph{p}($\omega$) is the prior density of the parameters $\omega$; \emph{x} and \emph{t} are input and output data; \emph{p}($\omega$$\mid$x,t) is the probability distribution of parameters $\omega$ after considering the data ($x,t$), i.e., the posterior distribution; \emph{p}($x,t$) is a normalization factor which ensures the integral of posterior distribution is one.

We adopt a Gaussian distribution for the likelihood based on an objective function, which is written as
\begin{equation}
    p(x,t\mid \omega)=\exp(-\chi ^{2}/2),
                                                                  \label{eqn.02}
\end{equation}
where the objective function $\chi$$^{2}$($\omega$) reads:
\begin{equation}
    \chi^{2}(\omega)=\sum^{N}_{i=1}(\frac{t_{i}-f(x_{i},\omega)}{\Delta t_{i}})^{2},
                                                                  \label{eqn.03}
\end{equation}
Here \emph{N} is the number of  data points, and $\Delta$$t_{i}$ is the associated noise scale which is related to specific observables. The function \emph{f}(x$_{i}$, $\omega$) depends on the input data x$_{i}$ and the model parameters $\omega$. In this work, the inputs of the network are given by x$_{i}$=\{Z$_{fi}$,Z$_{i}$,A$_{i}$,E$_{i}$\}, which include the charge number Z$_{fi}$ of the fission fragments, the charge number Z$_{i}$ and mass number A$_{i}$ of the fission nuclei and the excitation energy of the compound nucleus E$_{i}$=e$_{i}$+S$_{i}$(e$_{i}$ and S$_{i}$ are incident neutron energy and neutron separation energy, respectively); $t_{i}$ are the fission charge yields.

The posterior distributions are obtained by learning the given data. With new data $x_n$,
we make predictions by averaging the neural network over the posterior probability density of the network parameters $\omega$,
\begin{equation}
    \langle f_{n} \rangle=\int f(x_{n},\omega)p(\omega \mid x,t)d\omega,
                                                                  \label{eqn04}
\end{equation}
The high-dimensional integral in Eq.\ref{eqn04} is approximated by Monte Carlo integration in which the posterior probability \emph{p}($\omega$$\mid$$x, t$) is sampled using the Markov Chain Monte Carlo method~\cite{Neal1996}.

In BNN we need to specify the form of the functions \emph{f}(x, $\omega$) and \emph{p}($\omega$). In this work, we use a feed-forward neural network model defined the function \emph{f}(x, $\omega$). That is
\begin{equation}
    f(x,\omega)=a+\sum^{H}_{j=1}b_{j}\tanh(c_{j}+\sum^{I}_{i=1}d_{ji}x_{i}),
                                                                  \label{eqn.05}
\end{equation}
where \emph{H} is the number of neurons in the hidden layer, \emph{I} denotes the number of input variables and $\omega$=\{$a$, $b_{j}$, $c_{j}$, $d_{ji}$\} is the model parameters, $a$ is bias of output layers, $b_{j}$ are the weights of output layers, $c_{j}$ is bias of hidden layers, and $d_{ji}$ are weights of hidden layers. In total, the number of parameters in this neural network is 1+(2+\emph{I})$\times$\emph{H}.
To study the odd-even effects, an additional input variable to identify the odd-even charge number is employed.
The confidential interval (CI) at 95\% level is given for uncertainty quantifications in this work.
More details about  BNN  can be found in Ref.~\cite{Neal1996}.

\section{Results and discussions}
\begin{figure}[htbp]
\centering
\includegraphics[width=0.45\textwidth]{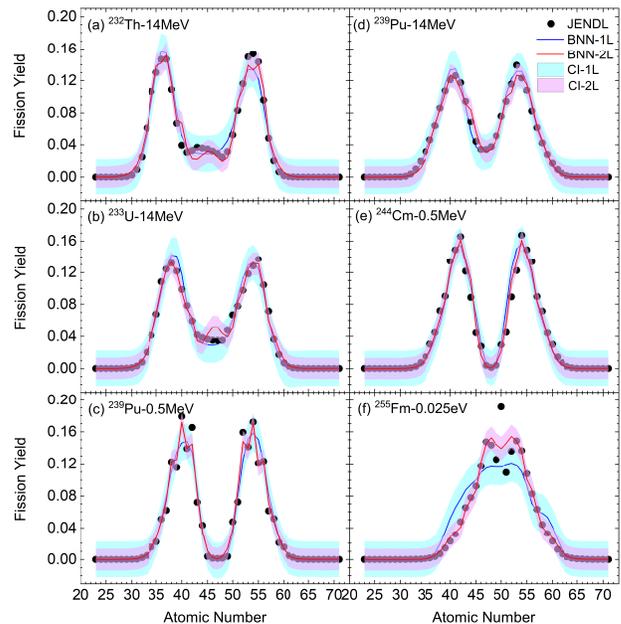}
\caption{
(Color online)
Comparison of one-layer (blue lines) and two-layer (red lines) BNN learning results of charge yields from JENDL~\cite{JENDL}. The shadow region corresponds to the confidence interval(CI) at 95\%.
                                                                   \label{FIG1}
}
\end{figure}

Firstly, we apply BNN to learn the existing evaluated distributions of charge yields from JENDL~\cite{JENDL}, which includes  2303 data points of
the neutron induced fissions of
 29 nuclei ($^{227,229,232}$Th, $^{231}$Pa, $^{232,233,234,236,237,238}$U,
$^{237,238}$Np, $^{238,239,240,241,242}$Pu, $^{241,243}$Am, $^{242,243,244,245,246,248}$Cm, $^{249, 251}$Cf, $^{254}$Es, $^{255}$Fm ).
We adopt a single hidden layer network of 32 neurons and a double hidden layer network of 16-16 neurons for comparison.
 We adopt 10$^5$ BNN sampling iterations in all calculations in this work.
A large number of sampling iterations are required for large data sets and large parameter sets.
 Note that the computing costs of BNN are very high compared to standard neural networks.
 The obtained standard deviations $\chi_N^2$=$\sum\limits_i (t_i-f(x_i))^2/N$
 are 1.43$\times$10$^{-5}$ and 6.36$\times$10$^{-6}$ for the single-layer and double-layer networks respectively. The single-layer network has been used previously for mass yields~\cite{fissionPRL2019}.
 Generally the double-layer network has much improved the learning performance
 of charge yields.  In this work, the charge distributions are trained independently and the normalization of the charge distributions
 are examined, which is close to the required 2.0 within 2\%.
 Fig.\ref{FIG1} displays the BNN learning results of charge distributions of 6 nuclei.
 It is shown that the single-layer network is less precise compared to the double-layer network.
 In particular, the single-layer network is not satisfactory for descriptions of $^{255}$Fm. Similarly, it was shown previously
 that the single-layer network is not satisfactory for mass yields around $^{227}$Th and $^{255}$Fm where neighboring nuclei in the learning set are not sufficient~\cite{fissionPRL2019}.
 The confidential interval (CI) at 95\% level are also shown in Fig.\ref{FIG1}.
 It is consistent that CI of the double-layer network is much smaller than that of the single-layer network.
 In $^{239}$Pu, the odd-even effects in charge yields are shown and the double-layer network can reproduce the peak structures while
 the results of the single-layer network are rather smooth.
 Note that the double-layer network has more connection parameters than that of the single-layer network, although they have the same number of neurons.
 There is also a risk of overfitting or training convergence issues with more parameters.
 Both the parameter numbers and the architecture can affect the network performance.
After above considerations, we choose the double-layer network which is suitable for the present study.

\begin{figure}[htbp]
\centering
\includegraphics[width=0.45\textwidth]{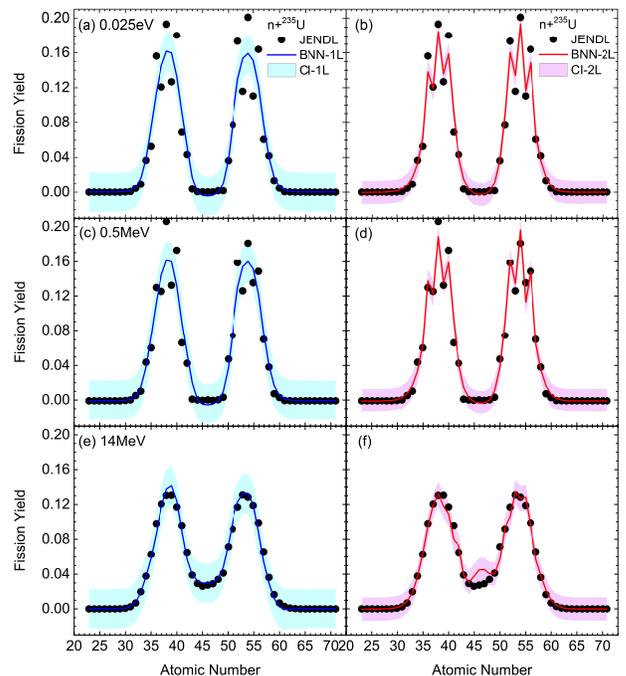}
\caption{
(Color online)
The BNN predicted fission  charge yields of n+$^{235}$U at neutron energies of 0.025 eV((a)(b)), 0.5 MeV((c)(d)) and 14 MeV((e)(f)), after learning the JENDL data without $^{235}$U.
The results of one-layer(left) and two-layer(right) are compared.  The shadow region corresponds to CI at 95\%.
                                                                   \label{FIG2}
}
\end{figure}


\begin{figure}[htbp]
\centering
\includegraphics[width=0.4\textwidth]{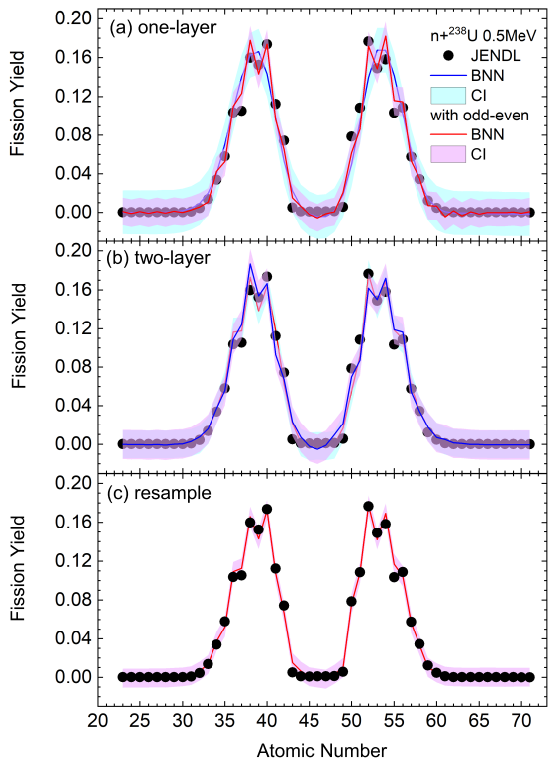}
\caption{
(Color online)
The BNN results of fission charge distribution of n+$^{238}$U at neutron energy of 0.5 MeV. (a) displays the comparison of  one-layer results without and with odd-even input. (b) displays the comparison of two-layer results. (c) displays results of with odd-even input and resampled learning.  The shadow region corresponds to CI at 95\%.
                                                               \label{FIG3}
}
\end{figure}

Next we test the predictive ability of BNN with a learning set without $^{235}$U,
compared with JENDL evaluation data.
The predicted charge distributions of $^{235}$U are shown in Fig.\ref{FIG2}.
It can be seen that single-layer results can not describe the detailed
peak structures of the charge distributions at low incident neutron energies.
At energies of 14 MeV, the single-layer results are better.
At low energies, the charge distributions have obvious odd-even effects, which
disappear at high excitations. We see the double-layer predictions can
describe the energy dependence of the odd-even effects.
It is also shown from CI that the double-layer predictions have smaller uncertainties than
that of the single-layer predictions.

It is known that odd-even effects are considerable in charge distributions while it is
not obvious in mass distributions. The charge yields of proton-even nuclei are larger than
that of proton-odd nuclei around peaks. To simulate the odd-even effects in BNN, we
add an additional input variable $\delta=\pm 0.2$ for indicating even and odd atomic numbers respectively.
The input parameter set now becomes x$_{i}$=\{Z$_{fi}$,Z$_{i}$,A$_{i}$,E$_{i}$, $\delta$\}.
Similar methods have been used in BNN to account odd-even pairing correlations and shell corrections
in estimations of global nuclear masses~\cite{Niu2018PLB}.
Fig.\ref{FIG3} displays the training performance of charge distributions of n+$^{238}$U at neutron energy of 0.5 MeV.
It can be seen that with the single-layer network, the influences of additional odd-even input $\delta$ is
significant, while the odd-even effects is not shown without the $\delta$ input. The associated CI with
$\delta$ input is slightly reduced.
For the double-layer network, the odd-even effects are shown with and without the $\delta$ input,
with similar uncertainties.
The additional $\delta$ input doesn't gain much performance for the double-layer network.
Our main motivation is to evaluate the charge distributions of the compound nucleus $^{239}$U,
so that we trained BNN with resampled learning of the evaluated n+$^{238}$U data from JENDL,
as shown in Fig.\ref{FIG3}(c). In this case, we reproduce learning data very precisely
with the double-layer network plus odd-even input and resampled learning of  n+$^{238}$U data.

\begin{figure}[htbp]
\centering
\includegraphics[width=0.4\textwidth]{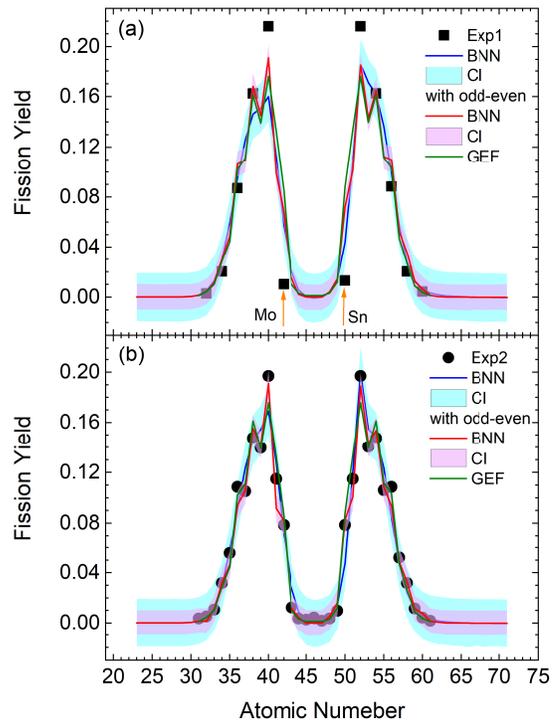}
\caption{
(Color online)
The BNN evaluations of fission charge yields of the compound nucleus $^{239}$U with two different experimental data. (a) the experiment corresponds to an excitation energy of 6.5MeV~\cite{PRL2017U}. (b) the experiment corresponds to an excitation energy of 8.3MeV~\cite{PRL2019U}. The blue lines and red lines are the two-layer BNN plus resampled learning results,  without and with odd-even input, respectively. The olive lines are the GEF evaluations taken from~\cite{PRL2019U}. The shadow region corresponds to CI at 95\%.
                                                                   \label{FIG4}
}
\end{figure}

Finally, we evaluate the charge distributions of fission fragments of the compound nucleus $^{239}$U, based on
the recently experimental data. In a recent experiment, Wilson \emph{et al.}~\cite{PRL2017U} for the first time used a novel technique which involves the coupling of a high-efficiency $\gamma$-ray spectrometer to an inverse-kinematics neutron source to extract charge yields of fission fragments via $\gamma$-$\gamma$ coincidence spectroscopy of $^{238}$U(n,f).
This experiment data is compared with results of the GEF evaluations~\cite{GEF} and charge yields around Sn and Mo isotopes are significantly small with a deviation by 600\%.
 However, in another recent experiment by  Ramos \emph{et al.}~\cite{PRL2019U}, direct measurements of isotopic fission yields of $^{239}$U performed using the neutron-transfer $^{9}$Be($^{238}$U,$^{239}$U)$^{8}$Be reaction don't show the abnormal deviation. The excitation energies in $^{239}$U in two experiments  are 6.5 MeV and 8.3 MeV respectively, which should have more or less similar fission yields.
The significant discrepancy can impact fission studies and nuclear applications.

Fig.\ref{FIG4} displays the BNN evaluations of the two experimental data. In Fig.\ref{FIG4}, BNN adopts the double-layer network with and without the odd-even indication.
The learning data set includes evaluated charge distributions from JENDL and the two incomplete experimental data.
In particular, the n+$^{238}$U data from JENDL has been resampled twice.
The evaluations without odd-even input and resamplings would not be satisfied.
It can be seen that the evaluations with odd-even input have significantly improved descriptions of the peak details.
The associated CI with odd-even input is also much smaller than that of evaluations without odd-even input.
The controversial charge yields around Sn and Mo isotopes in two evaluations are not small by our approach.
This is confirmed even without resampling n+$^{238}$U data.
This is consistent with the latest experiment that the abnormal deviation is not seen.
In addition, we speculate that the experimental charge yields at peaks around $Z$=40 and $Z$=52 could be too large based on BNN evaluations.
The BNN evaluation is also very successful for the 2017 data with clear odd-even effects, although which has very few data points of even atomic number.
The evaluations by GEF are also shown~\cite{PRL2019U}. For this particular case, BNN evaluations are comparable to
GEF evaluations.







\section{summary}

In summary, we applied the double-layer Bayesian neural network to learn and predict charge yields of fission fragments for the first time.
We found the performances of the double-layer network performance are significantly better than the single-layer network although they have the same number of neurons.
The double-layer network can describe the odd-even effects of charge yields at low energies, while odd-even effects are not obvious in mass yields.
We also add an additional input in BNN to indicate the odd-even atomic number which are very useful to improve evaluations.
We apply these methods to evaluate the incomplete charge yields of two recent experiments of $^{239}$U.
Our BNN evaluations don't obtain abnormal small charge yields around Sn and Mo isotopes as reported in Ref.~\cite{PRL2017U}.
This is consistent with the latest experiment~\cite{PRL2019U}.
The BNN evaluations are comparable to GEF evaluations for this particular evaluations.
Further improvement of BNN is still underway and is promising for quantitative modeling fission data for practical nuclear applications.

\acknowledgments
 This work was supported by  National Key R$\&$D Program of China (Contract No. 2018YFA0404403),
 and the National Natural Science Foundation of China under Grants No. 11975032, 11790325,11790320, 11835001, 11961141003.

\end{document}